\documentclass[final,3p,times,twocolumn,longtitle]{elsarticle}

\biboptions{compress}

\usepackage[draft]{hyperref}
\usepackage{amssymb}
\usepackage{color}
\usepackage{geometry}
\usepackage{epstopdf}

\journal{Astroparticle Physics} 

\begin{document}

\begin{frontmatter}

\title{On-orbit Operations and Offline Data Processing of CALET onboard the ISS}

\author[AF0,AF1]{Y.~Asaoka\corref{cor1}}
\author[AF0]{S.~Ozawa}
\author[AF0,AF1,AF2]{S.~Torii}
\author[AF3,AF4]{O.~Adriani}
\author[AF5,AF6]{Y.~Akaike}
\author[AF7]{K.~Asano}
\author[AF8,AF9]{M.G.~Bagliesi}
\author[AF8,AF9]{G.~Bigongiari}
\author[AF10]{W.R.~Binns}
\author[AF8,AF9]{S.~Bonechi}
\author[AF3,AF4]{M.~Bongi}
\author[AF8,AF9]{P.~Brogi}
\author[AF10]{J.H.~Buckley}
\author[AF12]{N.~Cannady}
\author[AF13]{G.~Castellini}
\author[AF14,AF15]{C.~Checchia}
\author[AF12]{M.L.~Cherry}
\author[AF14,AF15]{G.~Collazuol}
\author[AF16,AF17]{V.~Di~Felice}
\author[AF18]{K.~Ebisawa}
\author[AF18]{H.~Fuke}
\author[AF12]{T.G.~Guzik}
\author[AF5,AF19]{T.~Hams}
\author[AF20]{M.~Hareyama}
\author[AF21]{N.~Hasebe}
\author[AF22]{K.~Hibino}
\author[AF23]{M.~Ichimura}
\author[AF24]{K.~Ioka}
\author[AF7]{W.~Ishizaki}
\author[AF10]{M.H.~Israel}
\author[AF12]{A.~Javaid}
\author[AF21]{K.~Kasahara}
\author[AF21]{J.~Kataoka}
\author[AF25]{R.~Kataoka}
\author[AF26]{Y.~Katayose}
\author[AF27]{C.~Kato}
\author[AF28,AF29]{N.~Kawanaka}
\author[AF30]{Y.~Kawakubo}
\author[AF48]{K.~Kohri} 
\author[AF10]{H.S.~Krawczynski}
\author[AF19,AF5]{J.F.~Krizmanic}
\author[AF23]{S.~Kuramata}
\author[AF31,AF9]{T.~Lomtadze}
\author[AF8,AF9]{P.~Maestro}
\author[AF8,AF9]{P.S.~Marrocchesi}
\author[AF31,AF9]{A.M.~Messineo}
\author[AF6]{J.W.~Mitchell}
\author[AF32]{S.~Miyake}
\author[AF33]{K.~Mizutani\fnref{fn1}}
\author[AF34,AF19]{A.A.~Moiseev}
\author[AF21,AF18]{K.~Mori}
\author[AF35]{M.~Mori}
\author[AF4]{N.~Mori}
\author[AF36]{H.M.~Motz}
\author[AF27]{K.~Munakata}
\author[AF21]{H.~Murakami}
\author[AF37]{S.~Nakahira}
\author[AF18]{J.~Nishimura}
\author[AF38]{G.A.~de~Nolfo}
\author[AF22]{S.~Okuno}
\author[AF39]{J.F.~Ormes}
\author[AF3,AF13,AF4]{L.~Pacini}
\author[AF16,AF17]{F.~Palma}
\author[AF4]{P.~Papini}
\author[AF8,AF40]{A.V.~Penacchioni}
\author[AF10]{B.F.~Rauch}
\author[AF13,AF4]{S.B.~Ricciarini}
\author[AF19,AF5]{K.~Sakai}
\author[AF30]{T.~Sakamoto}
\author[AF19,AF34]{M.~Sasaki}
\author[AF22]{Y.~Shimizu}
\author[AF41]{A.~Shiomi}
\author[AF16,AF17]{R.~Sparvoli}
\author[AF3]{P.~Spillantini}
\author[AF8,AF9]{F.~Stolzi}
\author[AF42]{I.~Takahashi}
\author[AF18]{M.~Takayanagi}
\author[AF7]{M.~Takita}
\author[AF22]{T.~Tamura}
\author[AF22]{N.~Tateyama}
\author[AF37]{T.~Terasawa}
\author[AF18]{H.~Tomida}
\author[AF43]{Y.~Tsunesada}
\author[AF44]{Y.~Uchihori}
\author[AF18]{S.~Ueno}
\author[AF4]{E.~Vannuccini}
\author[AF12]{J.P.~Wefel}
\author[AF45]{K.~Yamaoka}
\author[AF46]{S.~Yanagita}
\author[AF30]{A.~Yoshida}
\author[AF47]{K.~Yoshida}
\author[AF7]{T.~Yuda\fnref{fn1}}
\address[AF0]{Waseda Research Institute for Science and Engineering, Waseda University, 3-4-1 Okubo, Shinjuku, Tokyo 169-8555, Japan}
\address[AF1]{JEM Utilization Center, Human Spaceflight Technology Directorate, Japan Aerospace Exploration Agency, 2-1-1 Sengen, Tsukuba, Ibaraki 305-8505, Japan}
\address[AF2]{School of Advanced Science and Engineering, Waseda University, 3-4-1 Okubo, Shinjuku, Tokyo 169-8555, Japan}
\address[AF3]{Department of Physics, University of Florence, Via Sansone, 1 - 50019 Sesto, Fiorentino, Italy}
\address[AF4]{INFN Sezione di Florence, Via Sansone, 1 - 50019 Sesto, Fiorentino, Italy}
\address[AF5]{Department of Physics, University of Maryland, Baltimore County, 1000 Hilltop Circle, Baltimore, MD 21250, USA}
\address[AF6]{Astroparticle Physics Laboratory, NASA/GSFC, Greenbelt, MD 20771, USA}
\address[AF7]{Institute for Cosmic Ray Research, The University of Tokyo, 5-1-5 Kashiwa-no-Ha, Kashiwa, Chiba 277-8582, Japan}
\address[AF8]{Department of Physical Sciences, Earth and Environment, University of Siena, via Roma 56, 53100 Siena, Italy}
\address[AF9]{INFN Sezione di Pisa, Polo Fibonacci, Largo B. Pontecorvo, 3 - 56127 Pisa, Italy}
\address[AF10]{Department of Physics, Washington University, One Brookings Drive, St. Louis, MO 63130-4899, USA}
\address[AF12]{Department of Physics and Astronomy, Louisiana State University, 202 Nicholson Hall, Baton Rouge, LA 70803, USA}
\address[AF13]{Institute of Applied Physics (IFAC),  National Research Council (CNR), Via Madonna del Piano, 10, 50019 Sesto, Fiorentino, Italy}
\address[AF14]{Department of Physics and Astronomy, University of Padova, Via Marzolo, 8, 35131 Padova, Italy}
\address[AF15]{INFN Sezione di Padova, Via Marzolo, 8, 35131 Padova, Italy}
\address[AF16]{University of Rome ``Tor Vergata'', Via della Ricerca Scientifica 1, 00133 Rome, Italy}
\address[AF17]{INFN Sezione di Rome ``Tor Vergata'', Via della Ricerca Scientifica 1, 00133 Rome, Italy}
\address[AF18]{Institute of Space and Astronautical Science, Japan Aerospace Exploration Agency, 3-1-1 Yoshinodai, Chuo, Sagamihara, Kanagawa 252-5210, Japan}
\address[AF19]{CRESST and Astroparticle Physics Laboratory NASA/GSFC, Greenbelt, MD 20771, USA}
\address[AF20]{St. Marianna University School of Medicine, 2-16-1, Sugao, Miyamae-ku, Kawasaki, Kanagawa 216-8511, Japan}
\address[AF21]{Research Institute for Science and Engineering, Waseda University, 3-4-1 Okubo, Shinjuku, Tokyo 169-8555, Japan}
\address[AF22]{Kanagawa University, 3-27-1 Rokkakubashi, Kanagawa, Yokohama, Kanagawa 221-8686, Japan}
\address[AF23]{Faculty of Science and Technology, Graduate School of Science and Technology, Hirosaki University, 3, Bunkyo, Hirosaki, Aomori 036-8561, Japan}
\address[AF24]{Yukawa Institute for Theoretical Physics, Kyoto University, Kitashirakawa Oiwakecho, Sakyo, Kyoto 606-8502, Japan}
\address[AF25]{National Institute of Polar Research, 10-3, Midori-cho, Tachikawa, Tokyo 190-8518, Japan}
\address[AF26]{Faculty of Engineering, Division of Intelligent Systems Engineering, Yokohama National University, 79-5 Tokiwadai, Hodogaya, Yokohama 240-8501, Japan}
\address[AF27]{Faculty of Science, Shinshu University, 3-1-1 Asahi, Matsumoto, Nagano 390-8621, Japan}
\address[AF28]{Hakubi Center, Kyoto University, Yoshida Honmachi, Sakyo-ku, Kyoto, 606-8501, Japan}
\address[AF29]{Department of Astronomy, Graduate School of Science, Kyoto University, Kitashirakawa Oiwake-cho, Sakyo-ku, Kyoto, 606-8502, Japan}
\address[AF30]{College of Science and Engineering, Department of Physics and Mathematics, Aoyama Gakuin University,  5-10-1 Fuchinobe, Chuo, Sagamihara, Kanagawa 252-5258, Japan}
\address[AF48]{Institute of Particle and Nuclear Studies, High Energy Accelerator Research Organization, 1-1 Oho, Tsukuba, Ibaraki, 305-0801, Japan} 
\address[AF31]{University of Pisa, Polo Fibonacci, Largo B. Pontecorvo, 3 - 56127 Pisa, Italy}
\address[AF32]{Department of Electrical and Electronic Systems Engineering, National Institute of Technology, Ibaraki College, 866 Nakane, Hitachinaka, Ibaraki 312-8508 Japan}
\address[AF33]{Saitama University, Shimo-Okubo 255, Sakura, Saitama, 338-8570, Japan}
\address[AF34]{Department of Astronomy, University of Maryland, College Park, Maryland 20742, USA }
\address[AF35]{Department of Physical Sciences, College of Science and Engineering, Ritsumeikan University, Shiga 525-8577, Japan}
\address[AF36]{International Center for Science and Engineering Programs, Waseda University, 3-4-1 Okubo, Shinjuku, Tokyo 169-8555, Japan}
\address[AF37]{RIKEN, 2-1 Hirosawa, Wako, Saitama 351-0198, Japan}
\address[AF38]{Heliospheric Physics Laboratory, NASA/GSFC, Greenbelt, MD 20771, USA}
\address[AF39]{Department of Physics and Astronomy, University of Denver, Physics Building, Room 211, 2112 East Wesley Ave., Denver, CO 80208-6900, USA}
\address[AF40]{ASI Science Data Center (ASDC), Via del Politecnico snc, 00133 Rome, Italy}
\address[AF41]{College of Industrial Technology, Nihon University, 1-2-1 Izumi, Narashino, Chiba 275-8575, Japan}
\address[AF42]{Kavli Institute for the Physics and Mathematics of the Universe, The University of Tokyo, 5-1-5 Kashiwanoha, Kashiwa, 277-8583, Japan}
\address[AF43]{Division of Mathematics and Physics, Graduate School of Science, Osaka City University, 3-3-138 Sugimoto, Sumiyoshi, Osaka 558-8585, Japan}
\address[AF44]{National Institutes for Quantum and Radiation Science and Technology, 4-9-1 Anagawa, Inage, Chiba 263-8555, JAPAN}
\address[AF45]{Nagoya University, Furo, Chikusa, Nagoya 464-8601, Japan}
\address[AF46]{College of Science, Ibaraki University, 2-1-1 Bunkyo, Mito, Ibaraki 310-8512, Japan}
\address[AF47]{Department of Electronic Information Systems, Shibaura Institute of Technology, 307 Fukasaku, Minuma, Saitama 337-8570, Japan}
\cortext[cor1]{Corresponding author. \\ E-mail address: yoichi.asaoka@aoni.waseda.jp (Y.Asaoka)}
\fntext[fn1]{Deceased.}

\begin{abstract}
The CALorimetric Electron Telescope (CALET), launched for installation on the International Space Station (ISS) in August, 2015, has been accumulating scientific data since October, 2015. CALET is intended to perform long-duration observations of high-energy cosmic rays onboard the ISS. 
CALET directly measures the cosmic-ray electron spectrum in the energy range of 1 GeV to 20 TeV with a 2\% energy resolution above 30~GeV. 
In addition, the instrument can measure the spectrum of gamma rays well into the TeV range, and the spectra of protons and nuclei up to a PeV. 

In order to operate the CALET onboard ISS, JAXA Ground Support Equipment (JAXA-GSE) and the Waseda CALET Operations Center (WCOC) have been established at JAXA and Waseda University, respectively. Scientific operations using CALET are planned at WCOC, taking into account orbital variations of geomagnetic rigidity cutoff. Scheduled command sequences are used to control the CALET observation modes on orbit. 
Calibration data acquisition by, for example, recording pedestal and penetrating particle events, a low-energy electron trigger mode operating at high geomagnetic latitude, a low-energy gamma-ray trigger mode operating at low geomagnetic latitude, and an ultra heavy trigger mode, are scheduled around the ISS orbit while maintaining maximum exposure to high-energy electrons and other high-energy shower events by always having the high-energy trigger mode active. 
The WCOC also prepares and distributes CALET flight data to collaborators in Italy and the United States.

As of August 31, 2017, the total observation time is 689 days with a live time fraction of the total time of $\sim$84\%. Nearly 450 million events are collected with a high-energy ($E>$10~GeV) trigger. In addition, calibration data acquisition and low-energy trigger modes, as well as an ultra-heavy trigger mode, are consistently scheduled around the ISS orbit.
By combining all operation modes with the excellent-quality on-orbit data collected thus far, it is expected that a five-year observation period will provide a wealth of new and interesting results.

\end{abstract}


\end{frontmatter}

\section{Introduction}
\label{intro}
\begin{figure*}[t!]
\begin{center}
\vspace*{-1cm}
\begin{minipage}{0.55\hsize}
\includegraphics[width=1.00\hsize]{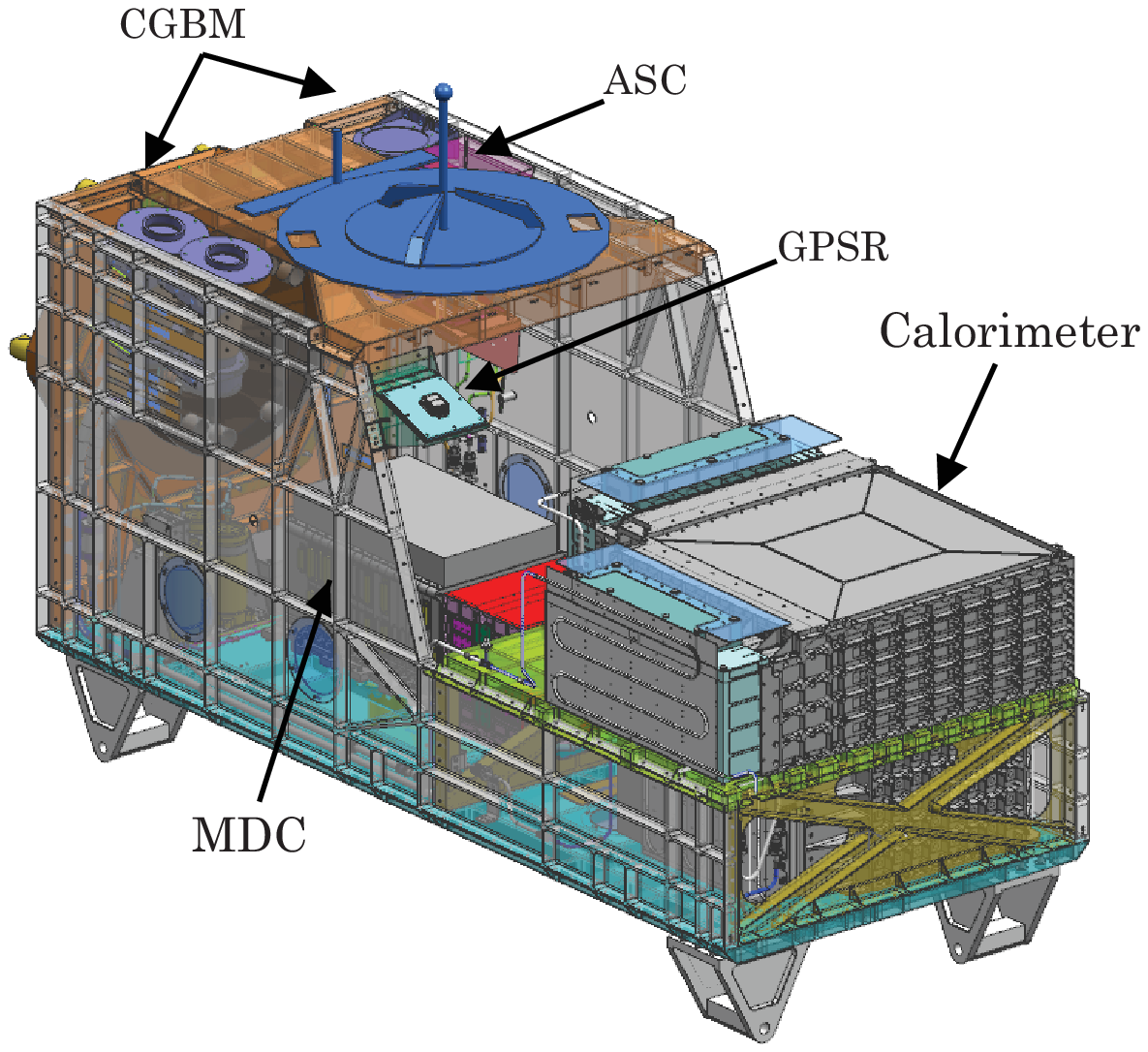}
\end{minipage}
\begin{minipage}{0.43\hsize}
\includegraphics[width=1.00\hsize]{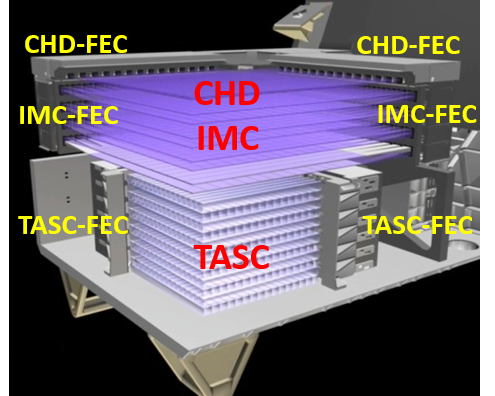}
\end{minipage}
\caption{({\it Left}) The CALET instrument package~\citep{torii2015} showing the main calorimeter, the  Gamma-ray Burst Monitor (CGBM) consisting of a hard X-ray monitor, and a soft gamma-ray monitor~\citep{CGBM2013}, Mission Data Controller (MDC) and support sensors, installed in a JEM standard payload with a size of 1850~mm (L) $\times$ 800~mm (W) $\times$ 1000~mm (H). 
Support sensors include Global Position Sensor Receiver (GPSR) and Advanced Stellar Compass (ASC) as indicated in the figure.
The total weight is 613~kg. 
({\it Right}) Layout of the main calorimeter, which consists of a Charge Detector (CHD), an IMaging Clorimeter (IMC), and Total AbSorption Calorimeter (TASC),
where FEC stands for front end electronics. }
\label{fig:calet}.
\end{center}
\end{figure*}

The CALorimetric Electron Telescope (CALET)~\citep{torii2015}, launched for installation on the International Space Station (ISS) in August, 2015, has been accumulating scientific data since October, 2015. CALET is primarily intended to discover nearby cosmic-ray accelerators and search for dark matter by precisely measuring all-electron (electron $+$ positron) and gamma-ray spectra in a wide energy range from 1~GeV to 20~TeV. CALET includes a high-performance particle detector equipped with a thick large-area calorimeter. Onboard the ISS, CALET has been performing long-term observations for two years and is expected be operational for three or more additional years.
A schematic overview of the CALET instrument is presented in the left-hand panel of Fig.~\ref{fig:calet}.

CALET features a very thick calorimeter that incorporates imaging and total absorption calorimeters (see the right-hand panel of Fig.~\ref{fig:calet}).
A calorimeter of 30 radiation-length thickness completely absorbs the electron shower energy in the TeV energy range and identifies electrons from the overwhelming flux of protons using the difference in shower development in the fully active fine-sampling and thick calorimeter.
Long-term observation using the large-area detector is provided by observation onboard the ISS.  
By combining all of these features, it becomes possible for the first time to precisely measure the all-electron spectrum up to 20 TeV.
The main components of cosmic rays, such as protons, heliums, 
and heavier nuclei, can be measured past PeV. 
Including electrons and gamma rays, the ability to perform unique observations by extending the previous limits of direct measurements is expected.

In this paper, we mainly describe the operations and offline data processing of the main calorimeter. 
Details of processing CGBM data (after creating Level-1 data: see Section~\ref{sec:dataproc})
are presented elsewhere~\citep{yamaoka2016,sergio2016}. 

\section{The CALET Detector System}
\subsection{Detector Components}
The CALET detector (see the right-hand panel of Fig.~\ref{fig:calet}) consists of a Charge Detector (CHD), which identifies the charge of the incident particle~\citep{pier2011,pier2013}, 
an IMaging Calorimeter (IMC), which reconstructs the track of the incident particle and finely images the initial shower development, and a Total AbSorption Calorimeter (TASC), which absorbs the entire energy of the incoming particle and identifies the particle species using hodoscopic scintillator arrays. 

The CHD is double layered (X,Y) and located above the IMC.
Each layer consists of 14 plastic scintillator paddles. 
Each paddle is read by the photomultiplier tube (PMT) and has the dimensions 450~mm (L) $\times$ 32~mm (W) $\times$10~mm (H).
The IMC 
consists of 7 layers of tungsten plates each separated by 2 layers of scintillating fiber (SciFi) belts 
arranged in the X and Y direction with an additional X,Y SciFi layer pair on the top. 
Each SciFi belt is assembled with 448 1~mm$^2$ cross-section scintillating fibers 
read by multi-anode photomultiplier tubes (MAPMT).
The dimensions of the SciFi layers are 448~mm (L) $\times$ 448~mm (W). 
The total thickness of the IMC is equivalent to 3 $X_0$. 
The first 5 tungsten-SciFi layers sample the shower every 0.2 $X_0$ and the last 2 layers provide 1.0 $X_0$ sampling.
The TASC is composed of 12 layers each of which consists of 16 lead tungstate (PWO) logs.
Alternate layers are arranged with the logs oriented along orthogonal (X,Y) directions. 
Each PWO log has dimensions of 326~mm (L) $\times$ 19~mm (W) $\times$ 20~mm (H).

Combining these components as well as the trigger system, data acquisition system and support sensors described in the following sections, the CALET detector features (1) a proton rejection factor of more than 10$^5$, (2) a 2\% energy resolution above 30~GeV, (3) an angular resolution of 0.1 to 0.5$^\circ$, and (4) a large geometrical factor on the order of 0.1~m$^2$sr. 
The basic features of detector performance were  investigated using Monte Carlo (MC) simulations~\citep{akaike2011}, and verified by the beam tests at CERN-SPS ~\citep{akaike2013, akaike2015, niita2015}.
The sufficiently high rejection capability of protons enables the suppression of  systematic errors in the electron spectrum due to 
uncertainties in proton rejection factor calculation due to possible interaction model
dependence. 
The CALET detector is the most suitable detector for directly measuring the all-electron spectrum up to 20~TeV.

\subsection{Trigger System}
\begin{figure*}[p!]
  \begin{center}
\includegraphics[width=0.88\hsize]{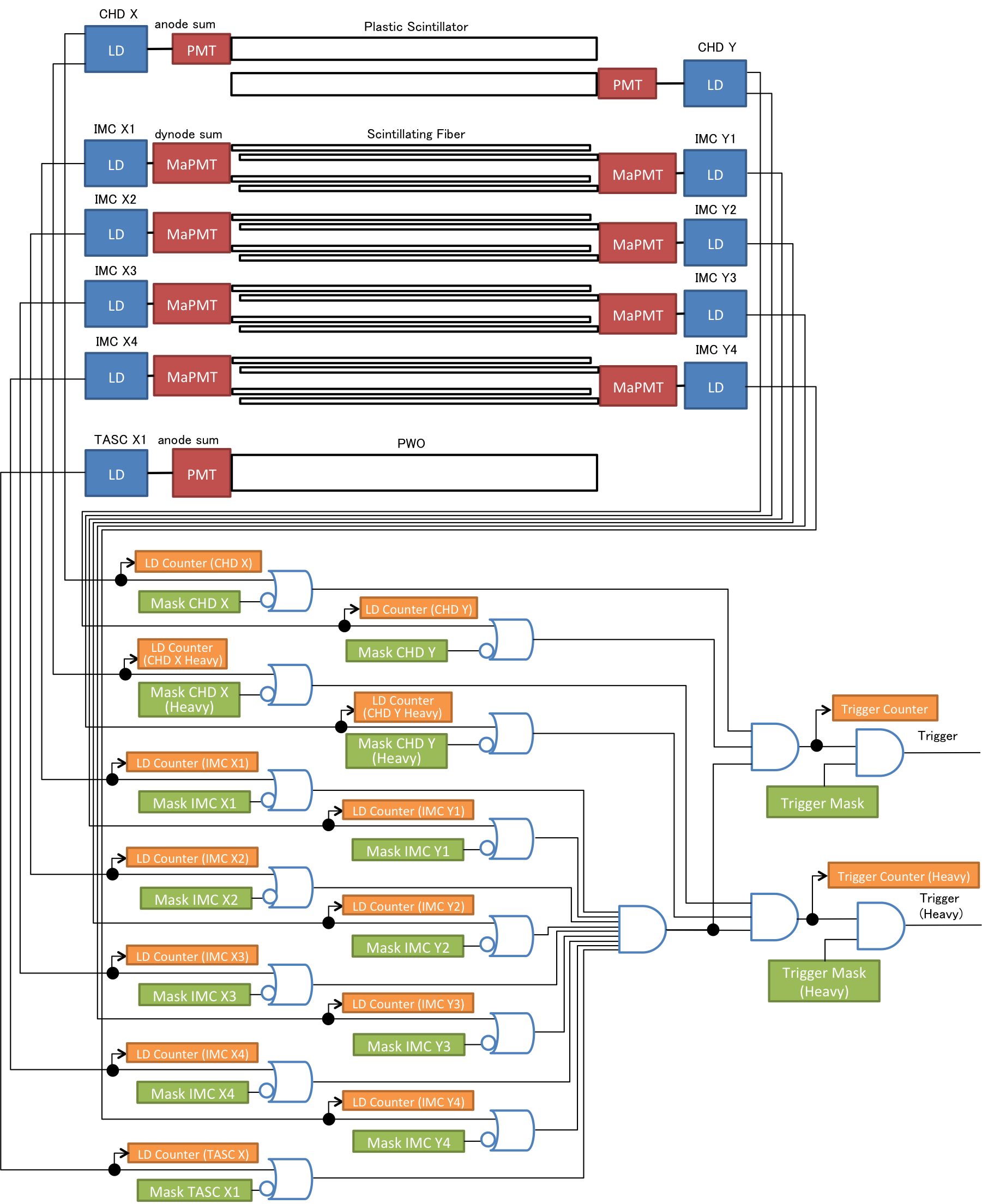}
  \end{center}
  \caption{Diagram of CALET's trigger logic for the LE trigger. The three trigger modes (Single, LE, and HE) use the same logic, and each trigger mode and its associated heavy mode is realized, as shown in the diagram. The dashed lines indicate the connections to the other two modes. The LD counters sum up the LD signals sent by each detector element and are read out when an event is built and stored together with the event data, or are read out in fixed intervals and stored as periodic data. The counters return to zero when reaching the maximum of the 24- or 32-bit buffer, where buffer size is determined taking into account the expected count rate for each counter. These counters are not part of the trigger logic but are used to monitor the trigger logic input.}
  \label{triggerLogic}
\end{figure*}

Since GeV cosmic rays are dominant in CALET, and it takes approximately 5 ms to be ready for the next event after building an event by retrieving all of the ADC data from the detector components, 
it is necessary to trigger CALET only for high energy particles.
CALET event selection is based on the coincidence of trigger counter signals generated from the detector discriminators, i.e., each of CHD X and Y, IMC X1-X4, Y1-Y4, and TASC X1 generates low level discriminator (LD) signals (see top of Fig.~\ref{triggerLogic}).
Since two IMC fiber layers along a given axis are read out in one front-end circuit, 
the number of trigger counter signals is reduced to four (from eight) in the X axis and four in Y axis. 
TASC X1 is read out by photomultipliers to retrieve fast trigger signals.
The other 11 TASC layers are read out by a photodiode (PD) and an avalanche photodiode (APD) in order to ensure a wide dynamic range of six orders of magnitude.

In order to efficiently collect data under different conditions, CALET features three trigger modes:
\begin{description}
 \item[{\bf High-energy Shower Trigger (HE): }] This is the main trigger mode for CALET 
	because it targets high-energy electrons of $10~{\rm GeV} \sim 20~{\rm TeV}$,
	high-energy gamma-rays of $10~{\rm GeV} \sim 10~{\rm TeV}$, and protons and nuclei
	of a few $10~{\rm GeV} \sim 1,000~{\rm TeV}$. By requiring a large energy deposit
	in the middle of the detector, high-energy shower events are selectively triggered.
	As a result, it is possible to maximize exposure to the target particles
	while strongly suppressing low-energy particles and achieving a large geometrical factor
	at the same time.
 \item[{\bf Low-energy Shower Trigger (LE): }]  This is the same as the high-energy shower trigger, but
	it targets lower-energy particles generating a shower in the detector. 
 	The primary targets of the LE trigger are
	low-energy electrons of $1~{\rm GeV} \sim 10~{\rm GeV}$  at high latitude,
	low-energy gamma-rays of $\geq 1~{\rm GeV}$ at low geomagnetic latitude, and
	GRB gamma rays of $\geq 1~{\rm GeV}$. In order to trigger electrons at high latitude,
	LD signals are required at CHD and IMC upper layers in order to restrict the incident angle of 
	the particles. The energy deposit required in the middle of the detector 
	is much lower than that for the HE trigger but is significantly 
	higher than that of a minimum ionizing particle (MIP).
 \item[{\bf Single Trigger (Single): }] This trigger mode is dedicated to taking data of non-interacting 
	particles for the detector calibration. This trigger mode requires the energy deposit of a MIP and 
	corresponding LD signals in the CHD and IMC, as well as the upmost layer of the TASC.
\end{description}
Since the combination of LDs required to generate the trigger signal is selectable,
other trigger modes, such as one dedicated to ultra-heavy nuclei~\citep{brian2011}, are also possible.

Trigger modes are realized as logical AND of LD signals as shown in Fig. \ref{triggerLogic}.
The anode sum signals of CHD X and Y, the dynode sum signals of IMC X1--X4 and Y1--Y4 layers, and the anode sum signals of TASC X1 are used as LD inputs. 
The front-end electronics have the capability to inhibit individual PMT signals at the input side of the analog summation. 
This allows the analog sum to be reconfigured in the event of a PMT malfunction.
The LD signal is generated when the input signal exceeds the discriminator threshold level.
As shown in Fig. \ref{triggerLogic}, 
the decision for a trigger mode is made by requiring coincident LD signals in a particular pattern.
The CHD has an additional LD, referred to as the heavy LD, which can be used to tag heavier nuclei. The heavy LD is used instead of the normal LD in the case of the heavy ion trigger mode, which is defined for each of the normal HE, LE, and Single triggers, resulting in six independent trigger modes. 

For each trigger mode, a certain set of LD signals is required for the trigger coincidence. 
The LD signals forming the trigger are selectable through use of a LD-mask pattern.
Setting the mask parameter of a particular LD to 1 (0) means that the corresponding LD signal is required (not required) to generate the trigger signal. 
Patterns of LD-mask are used to select different trigger modes. 
The IMC and TASC LD thresholds and the LD-mask values are common for both the heavy and normal modes.
For the HE, LE, and Single modes the CHD LD thresholds are common while the CHD LD mask values are different. 

\begin{figure}[hbtp]
  \begin{center}
   \includegraphics[width=1.0\hsize]{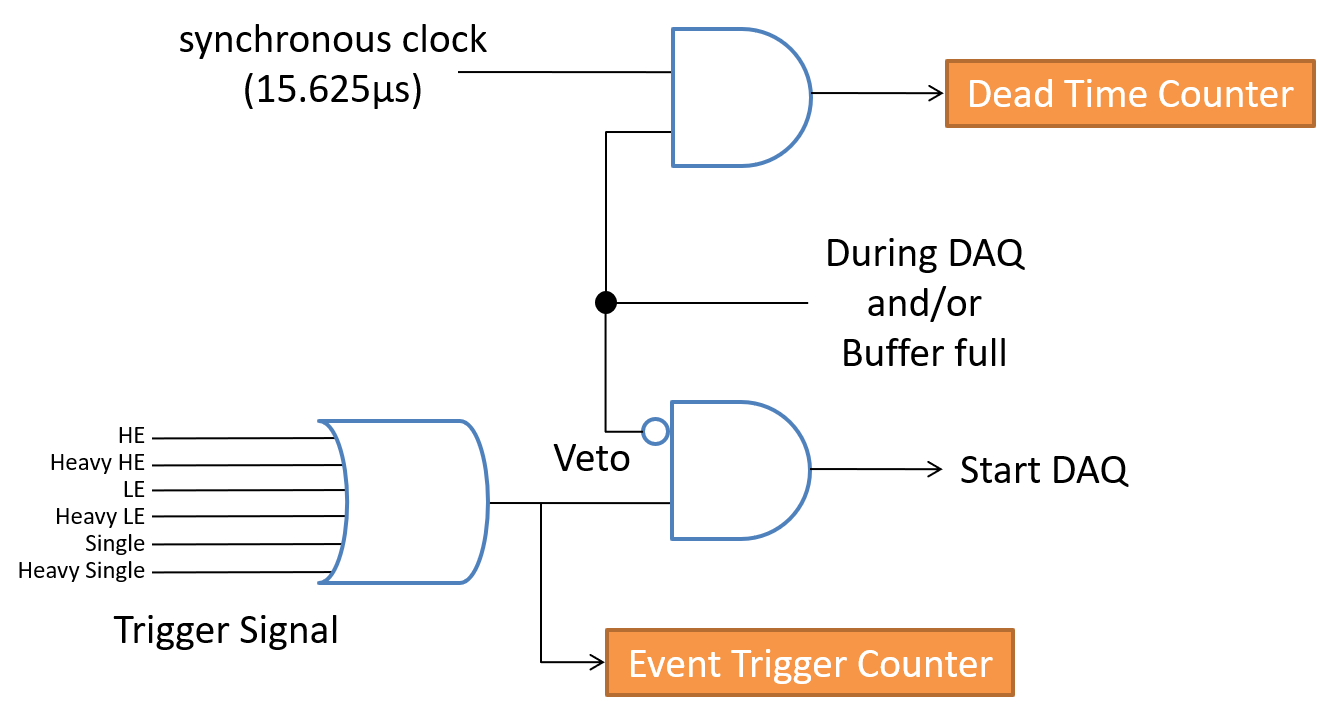}
  \end{center}
  \caption{Logic diagram of the trigger decision and data acquisition request for CALET.}
  \label{triggerDAQ}
\end{figure}

Figure~\ref{triggerDAQ} shows the logic diagram of the trigger decision. The trigger decision is performed by logical OR of the trigger signals from all of the 
trigger modes. Each event is flagged with all active modes that would have triggered the event, independent of which mode actually initiates the readout. Each trigger mode can be enabled/disabled by setting the trigger mask parameter. Only trigger modes with a trigger mask setting of 1 may generate an event trigger. 

Triggers are also inhibited if the DAQ is still building the previous events (dead time) or while all the Mission Data Controller (MDC) buffers are full.  
The buffers in the MDC are used to limit the CALET data downlink rate to a maximum of 600~kbps. As long as the buffers are not full, an increase in the data rate does not affect the data collection. 
The dead time counter counts the intervals of the 15.625 $\mu$s period clock, while triggers are inhibited in order to calculate the dead time fraction of the CALET observation. 

\subsection{Data Acquisition System}
The acquisition of cosmic-ray event data is carried out using the Mission Data Controller (MDC).
Event data acquisition consists of an event building task, an event processing task, and 
an event-delivering task. Since these tasks use the data buffer to communicate with each other,
it is possible to parallelize the tasks. 
Because of this design, the MDC can handle a sudden data increase, while limiting the observation dead time to the duration of the event building task,
thereby maximizing the observation live time.
Data downlink is controlled by the event delivering task and limits the bandwidth usage to a maximum of 600 kbps. 
As a complete readout of CALET exceeds 8000 channels, 
zero-suppression of the obtained ADC data is carried out in the event processing task
to compress the downlinked event.

In addition, the MDC handles control of the CALET detector components, 
the HV-Box which supplies high voltage to the detectors, 
the CALET Gamma-ray Burst Monitor (CGBM), and the support sensors. 
The support sensors include the Advanced Stellar Compass, which 
determines the attitude of CALET, and a GPS receiver, which 
is used during ground processing to correct the MDC time to the GPS time.
The MDC also handles data collection for the detectors and equipment. 
In the ever-changing geomagnetic radiation environment at the 
ISS orbit, 
it is necessary to 
control the trigger rate and data amount by 
programming the MDC with an appropriate schedule of 
trigger condition and 
zero-suppression threshold setting.
Planning such a schedule of scientific observations in order to effectively obtain useful data is one of the most important roles of WCOC.
The schedule also includes CGBM operations to activate and deactivate CGBM instruments.

\subsection{Support Sensors}
The absolute time scale is calibrated using a Global Position Sensor Receiver (GPSR) by
time stamping the pulse-per-second signal from the GPSR with the MDC time. 
The GPSR works as expected during nominal operations, and its duty cycle, i.e.,
the ratio of the time when the time pair is available to the total time, is 99.6\%.
The measured time drift of internal clock averaged over one year was determined as -1.6~s/day.
By correcting the drift using the most recently available time pair,
the UTC time is determined for each event.

The precise pointing direction is obtained using an attached Advanced Stellar Camera (ASC)~\citep{ASC}.
The ASC determines CALET's attitude using star images captured every half second.
Two methods are used to complement the data when the ASC attitude is not available because of the Sun:
\begin{enumerate}
\item Spherical-line interpolation (if the duration $<$1,000~s; it corresponds to 50\% of total observation time);
\item the use of corrected ISS quaternion (if the duration $>$1,000 s; it corresponds to 12\% of total observation time).
\end{enumerate}
After calibration using the time-dependent correction quaternion,
the discrepancy between the ASC and ISS quaternions is within 0.2~deg, indicating
valid attitude determination by the ASC considering the ISS quaternion accuracy.

\section{Ground System}
\begin{figure*}[tbh!]
\begin{center}
\includegraphics[bb=0 0 2233 698, width=1.0\hsize]{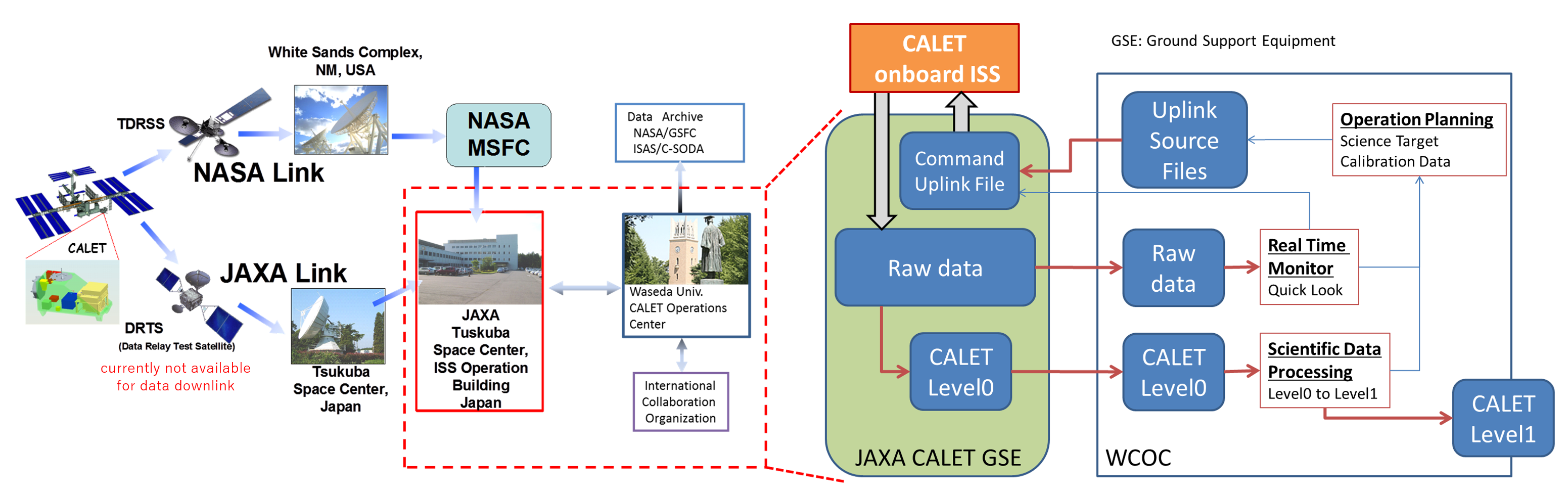}
\caption{CALET data flow from orbit to ground. The dataflow of the CALET ground system is summarized on the right-hand side. Interfaces to JAXA-GSE are defined corresponding to each role of WCOC. JAXA-GSE has direct interfaces to the ISS and CALET.
}
\label{fig:dataflow}
\end{center}
\end{figure*}

As shown in the left-hand panel of Fig.~\ref{fig:dataflow},
the data obtained from CALET onboard the ISS is transferred, 
through various satellite and ground links, to JAXA.
The JAXA Ground Support Equipment (JAXA-GSE) at JAXA and the Waseda CALET Operations Center (WCOC) at Waseda University were established in order to operate and monitor the CALET onboard the ISS. 
The raw data received by the JAXA-GSE are immediately transferred to the WCOC  for real-time instrument monitoring except for the replayed data stored at the ISS during the loss of signal.
Scientific raw data (Level~0) are also transferred from JAXA to WCOC on an hourly basis after the data records are time-order corrected and augmented, if necessary, with data replayed from ISS storage. 
In this context, the primary responsibility of the WCOC is to manage CALET scientific mission operations including 
(1) real-time monitoring and operations, (2) operations planning, and (3) scientific data processing. As shown in the right-hand panel of  Fig.~\ref{fig:dataflow}, there is an interface to a JAXA ground system corresponding to each role of WCOC, and providing uplink and downlink 
 communication with CALET onboard the ISS.

In order to monitor the observation status of CALET in real time, a quick look (QL) system, which consolidates and visualizes cosmic-ray event data and housekeeping data, was developed. Since a large amount of data must be monitored in real-time in a comprehensive manner, it is necessary to summarize the data and to detect malfunctions automatically. The QL system provides such functions. Using simulated CALET telemetry data, the QL systems are developed early on in a simulation of on-orbit operation, enabling the QL system to be used in CALET system tests, and immediately after its installation on the ISS. 
Currently, in the stable operation phase, using the QL system, the CALET science team monitors the scientific status and the data transmission on a 24-hour seven-day basis in WCOC.

\section{Data Flow and Data Analysis Framework} \label{sec:dataproc}
The scientific raw data (called CALET Level-0 data) are generated at JAXA and are transferred to the WCOC. 
The Level-0 data is, in essence, the raw CALET data stream with time ordered data packets collected from near real-time transmission or from replay of data stored on the ISS.
At the WCOC, the Level-0 data is converted into Level-1 data, 
and it is the Level-1 that is distributed to the international collaboration as the base data for the CALET scientific data analysis.
Quick analyses based on both Level-0 and Level-1 data are performed and their results are used as feedback for better operations planning and real-time monitoring. Level-1 data include all event, housekeeping, rate, and ancillary data records with timestamps corrected to UTC. The housekeeping temperatures, voltages, and currents are converted to physical units. 
\begin{figure}[htb]
\centering
\includegraphics[width=1.00\hsize]{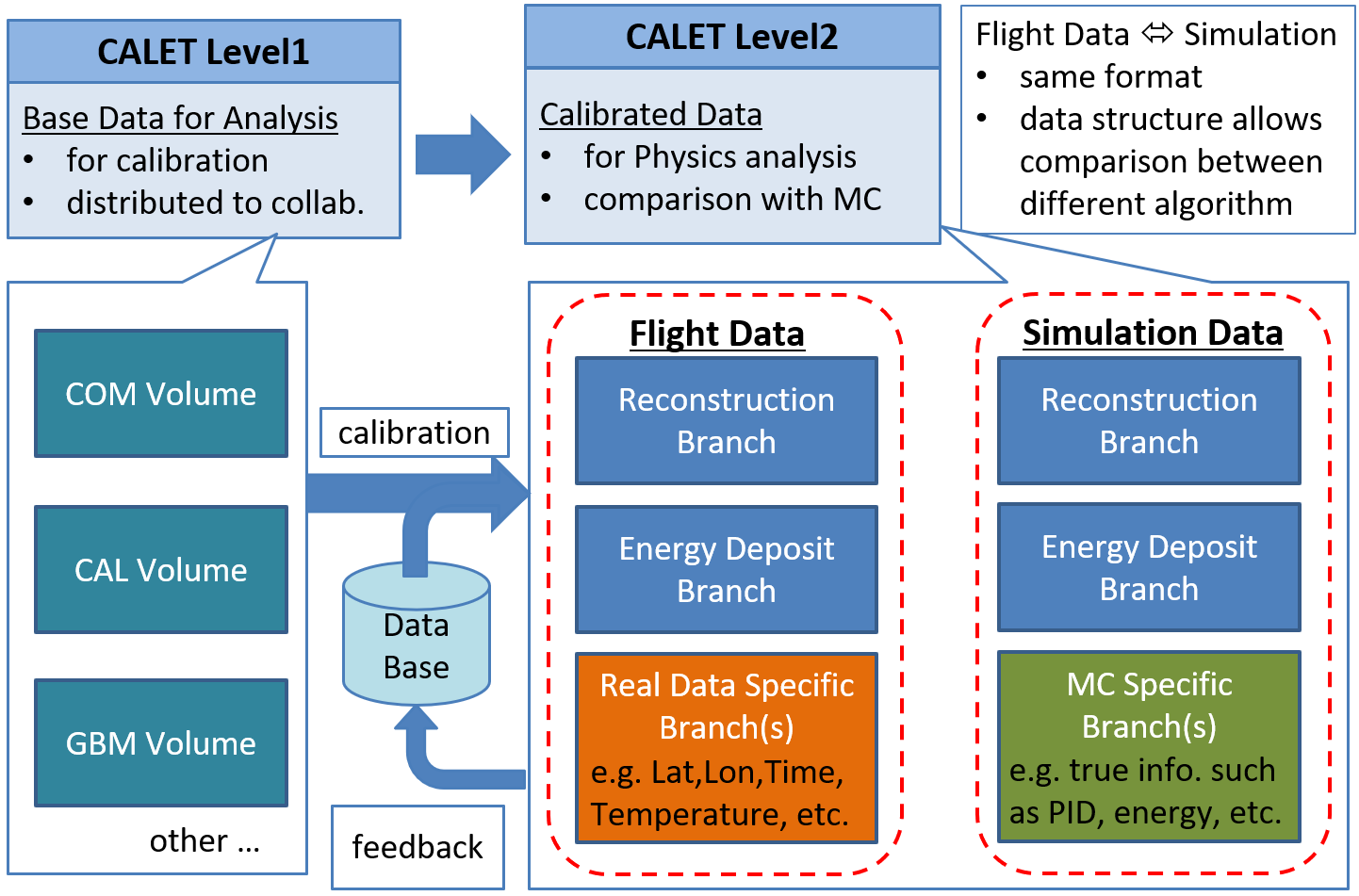}
\caption{Data analysis flow for creating calibrated data (Level-2). The calibration parameters are stored in a database and are used for iterative improvements of calibration, event reconstruction, and MC data.}
\label{dataprocess}
\end{figure}

The analysis scheme of CALET is shown in Figure \ref{dataprocess}. 
Various detector calibrations are performed using the Level-1 data and 
these calibrations are incorporated into the Level-2 data that is used for physics analysis~\citep{guzik2013,asaoka2017}.
For this physics analysis flight data (FD) and Monte Carlo simulation data (MC) are both in Level-2 format and are treated equivalently during analysis.
In order to produce the Level-2 data set, the energy deposits of all of the channels
are first calculated by applying detector calibrations to the FD and 
by smearing MC using the measured detector responses. 
In the process of Level-2 data production, the energy deposit array for both FD and MC is fed into track reconstruction~\citep{akaike2011, paolo2017} and energy reconstruction~\citep{asaoka2017, CALET2017-electron}). The various variables for event selection~(\citep{palma2015,pacini2017,pier2017,akaike2017} are calculated for both FD and MC. 
The TObjectArray in the ROOT analysis framework~\citep{root} provides a flexible format in which it is possible to record the results of multiple algorithms in the Level-2 data, allowing easy comparisons between different algorithms. 
By taking advantage of such Level-2 data features, an efficient and detailed study of systematic uncertainty becomes feasible~\citep{CALET2017-electron}.

The calculation nodes at WCOC have over 800 cores working at present, processing the high-level flight data and creating MC data. 
Spectral data and subsets of data for individual science targets are called Level-3 (or higher) data. Data analysis using each data set will be carried out independently in each institute, and the official dataset to be used for publication is processed at WCOC.

\section{On-orbit Operations}
\subsection{Observation Planning}
The CALET observation mode on orbit is controlled by a regular schedule of command sequences. 
A calibration data trigger mode that involves, for example, recording pedestal and penetrating particle events, a low-energy electron trigger mode operating at high geomagnetic latitude, and other dedicated trigger modes are scheduled around the ISS orbit while maintaining the maximum exposure to high-energy electrons.
\begin{figure}[tb!]
\begin{center}
\includegraphics[width=1.0\hsize]{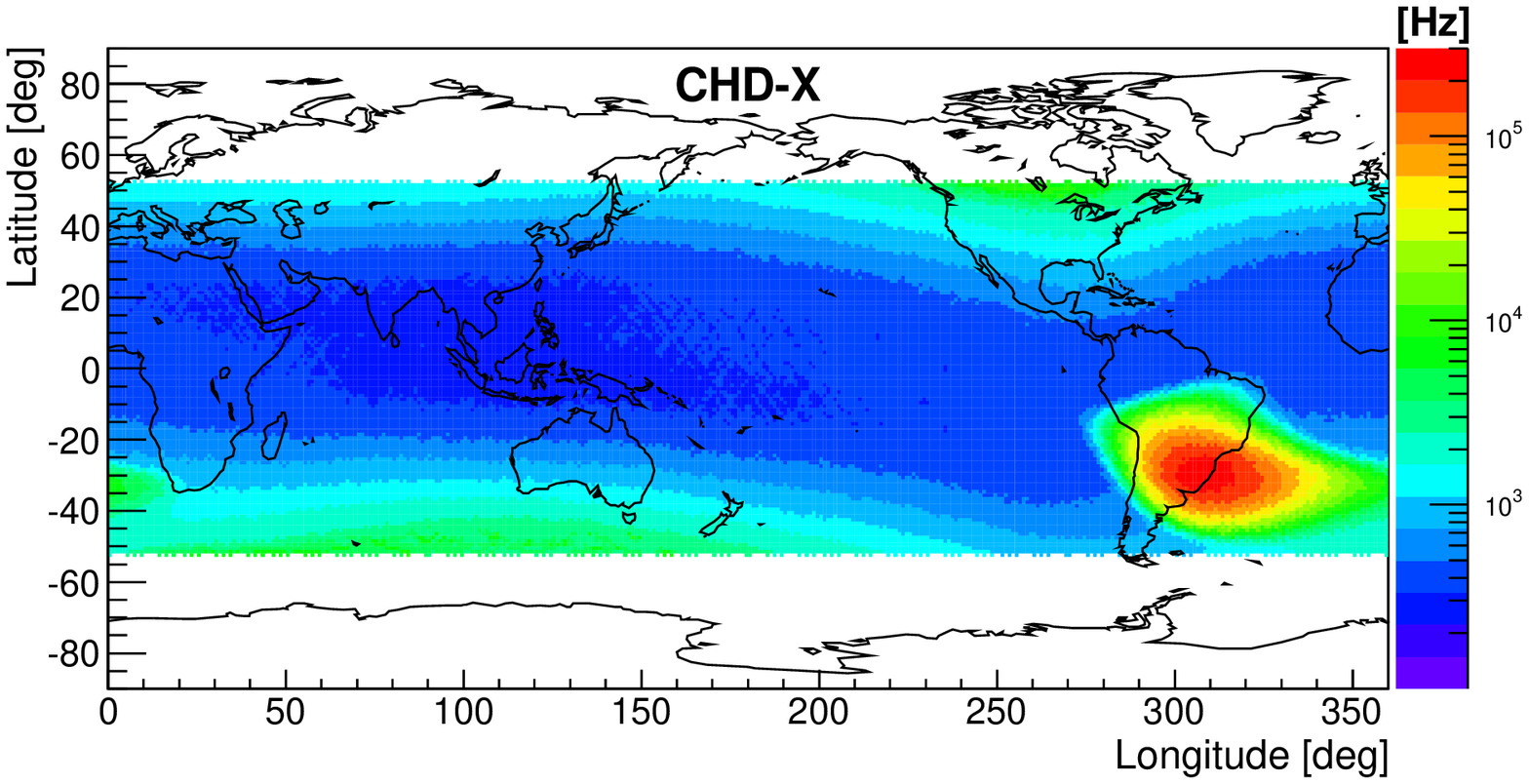}
\includegraphics[width=1.0\hsize]{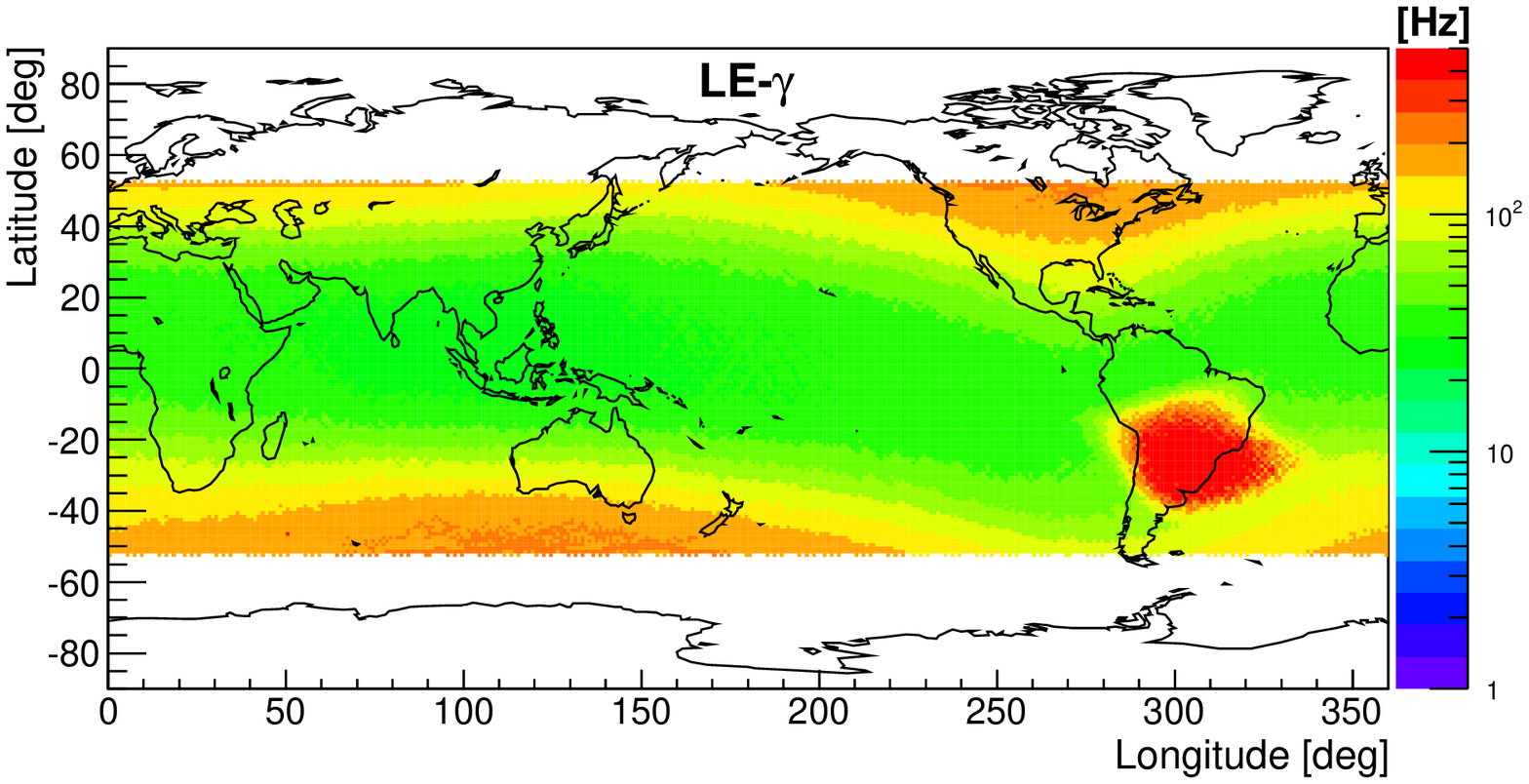}
\includegraphics[width=1.0\hsize]{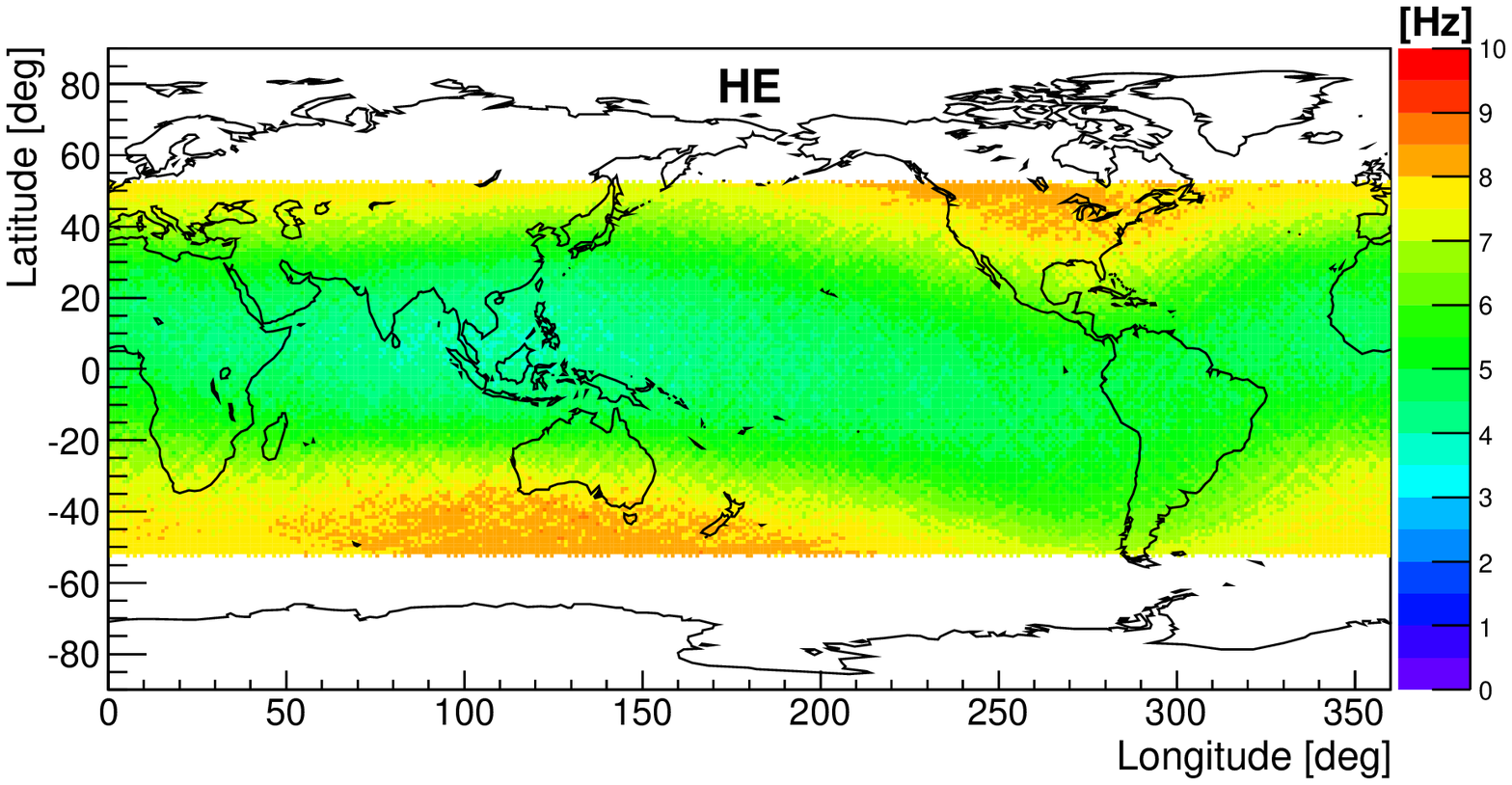}
\includegraphics[width=1.0\hsize]{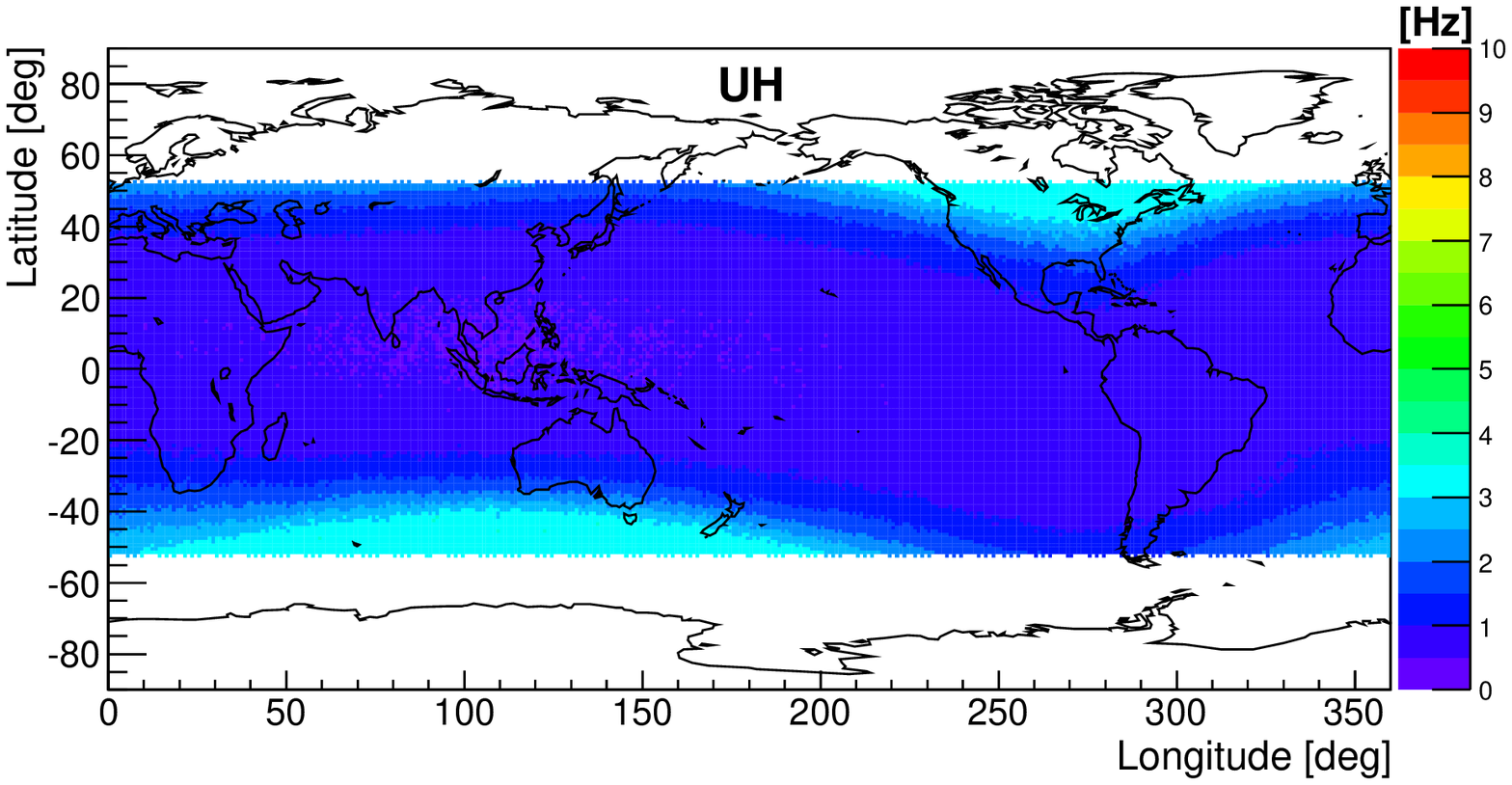}
\caption{Trigger/count rate dependence on the ISS position.
From top to bottom, the CHD-X count rate, LE-$\gamma$ trigger rate, 
HE trigger rate, and UH trigger rate are shown as color maps. 	
While the LE-$e$ trigger is selected at the highest geomagnetic latitude,
the maximum trigger rate is below 100~Hz, because of the requirements of
LD hits in the upper detector layers.
Note that the rate range in the color map is selected for each trigger mode so that
the dependence on the geomagnetic latitude is clear. 
}
\label{fig:rate}
\end{center}
\end{figure}

The following modes are combined in a schedule command file, and the trigger rate maps of representative observation modes, as well as a CHD count rate map, are shown in Fig~.\ref{fig:rate}. The LD threshold settings for each observation mode are summarized in Table~\ref{tab:thd}.
\begin{table}[hbt!] 
\caption{Examples of CALET trigger modes. For clarity, the LD thresholds, which are set as 8-bit digitized values, are shown in units of MIP in the table. In the table, ``-'' indicates that the corresponding LD signal is not required for the trigger.}
\label{tab:thd}
\begin{center}
\begin{scriptsize}
 \begin{tabular}{ccccccc} 
\hline\hline
&\\[-3pt]
     & CHD & IMC & IMC& IMC & IMC & TASC \\
Obs. & X & X1 & X2 & X3 & X4 & X1 \\
Mode & Y & Y1 & Y2 & Y3 & Y4 &    \\
     & [MIP] & [MIP]  & [MIP]  & [MIP]  & [MIP]  & [MIP] \\
\hline\hline
&\\[-3pt]
{\bf HE} & - & - & - & - & 16 & 76 \\ 
         & - & - & - & - & 16 &    \\
\hline
&\\[-3pt]
{\bf LE-$e$} 
         & 0.3 & 0.3 & 0.3 & 0.3 & 2.5 & 7 \\
         & 0.3 & 0.3 & 0.3 & 0.3 & 2.5 &   \\
\hline
&\\[-3pt]
{\bf LE-$\gamma$} 
         & - & - & - & - & 2.5 & 7 \\
         & - & - & - & - & 2.5 &   \\
\hline
&\\[-3pt]
{\bf S-$p$} 
         & 0.3 & 0.3 & 0.3 & 0.3 & 0.3 & 0.3 \\
         & 0.3 & 0.3 & 0.3 & 0.3 & 0.3 &     \\
\hline
&\\[-3pt]
{\bf S-He} 
         & 0.3 & 2.0 & 2.0 & 2.0 & 2.0 & 2.0 \\
         & 0.3 & 2.0 & 2.0 & 2.0 & 2.0 &     \\
\hline
&\\[-3pt]
{\bf UH} 
         & 100 & 40 & 40 & - & - & - \\ 
         & 100 & 40 & 40 & - & - &   \\ 
\hline\hline
 \end{tabular} 
\end{scriptsize}
\end{center}
\end{table}
\begin{itemize}
\item High-energy shower observation (HE): \\
All electrons and high-energy shower phenomena of gamma rays and nuclei are acquired.
The high-energy shower observation is always activated because this is the trigger mode for
the main objectives of CALET. Because of the high threshold, the South Atlantic Anomaly (SAA) does not have a significant impact on the trigger rate, as shown in the third panel from the top in Fig.~\ref{fig:rate}, making it possible to continue observation even in the SAA.
\item Low-energy electron observation  (LE-$e$): \\
For electron data in the 1-GeV region, this mode can acquire the low-energy data efficiently 
when the geomagnetic cutoff is low.
This mode is activated for 90~s at the highest geomagnetic latitude in the north and south regions.
\item Low-energy gamma-ray observation (LE-$\gamma$): \\
Using the geomagnetic cutoff for charged particles, low-energy gamma ray data is acquired efficiently. Low-energy gamma-ray observation is activated in the low-geomagnetic-latitude region when the geomagnetic latitude is below 20$^\circ$, except for the SAA.
Since it is possible to measure trigger rate without actually taking data, LE-$\gamma$ trigger rate for the whole orbit is shown in the second panel in Fig.~\ref{fig:rate}. 
\item MIP calibration data acquisition (S-$p$, S-He): \\
In order to check the equipment gain and stability, we collect non-interacting proton/helium events selectively.
MIP calibration data acquisition is normally activated during two consecutive orbits (three hours) per day in order to collect enough statistics to monitor the gain and detector stability.
\item Ultra-heavy nuclei observation (UH): \\
A dedicated trigger mode to acquire ultra-heavy nuclei penetrating the CHD
and the upper four layers of the IMC is implemented.
While the main target of the trigger mode is nuclei of Z$>$26,
the trigger threshold is loose enough to trigger Z$\geq$12 nuclei.
The ultra-heavy observation is almost always active because of the low
trigger rate.
\item Pedestal data acquisition:\\
Pedestal data are periodically acquired at a rate of 100 events every 23 minutes.
\end{itemize}
Since the UH trigger setting requires implementation as the LE trigger, LE-$e$ and LE-$\gamma$ observations use Single trigger in the actual observation. As a result, it is not possible to combine LE-$e$, LE-$\gamma$, S-$p$ and S-He observations at the same time.

The command file also includes CGBM control commands to protect the detector from high radiation at high geomagnetic latitude and in the SAA.
The execution time of the commands in the file take into account the ISS position and the Acquisition and Loss of Signal (AOS/LOS). 
The command schedule is created making use of recently observed data and is renewed every day semi-automatically in WCOC.

\subsection{Statistics of Accumulated Data}
\begin{figure}[htb]
\begin{center}
\includegraphics[width=1.00\hsize]{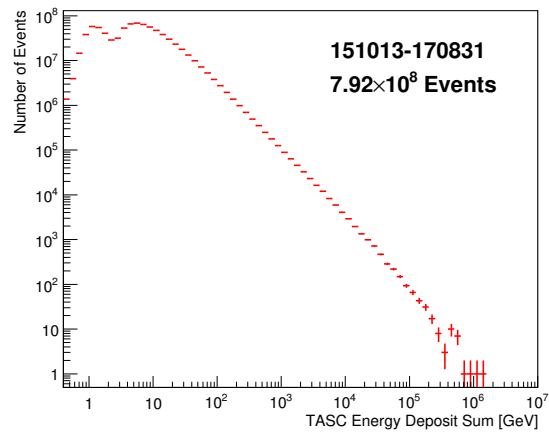}
\caption{TASC energy deposit spectrum using all of the triggered events through the end of August, 2017.}
\label{fig:edep}
\end{center}
\end{figure}

From the start of scientific observation, the accumulated number of events through the end of August 2017 is 7.92$\times$10$^8$.
Figure~\ref{fig:edep} shows the TASC energy deposit spectrum for the same period using all of the triggered events. 
The first bump is due to low-energy triggered events, while the second bump is caused by high-energy triggered events and the tail at high energy reflects the power-law nature of the cosmic-ray spectrum. 
The spectrum spans more than six orders of magnitude in energy with highest energy past a PeV, and the lowest energy below 1~GeV. 
This clearly demonstrates the CALET capability to observe cosmic rays over a very wide dynamic range.

\begin{figure*}[htb]
\centering
\begin{minipage}{0.49\hsize}
\includegraphics[width=1.00\hsize]{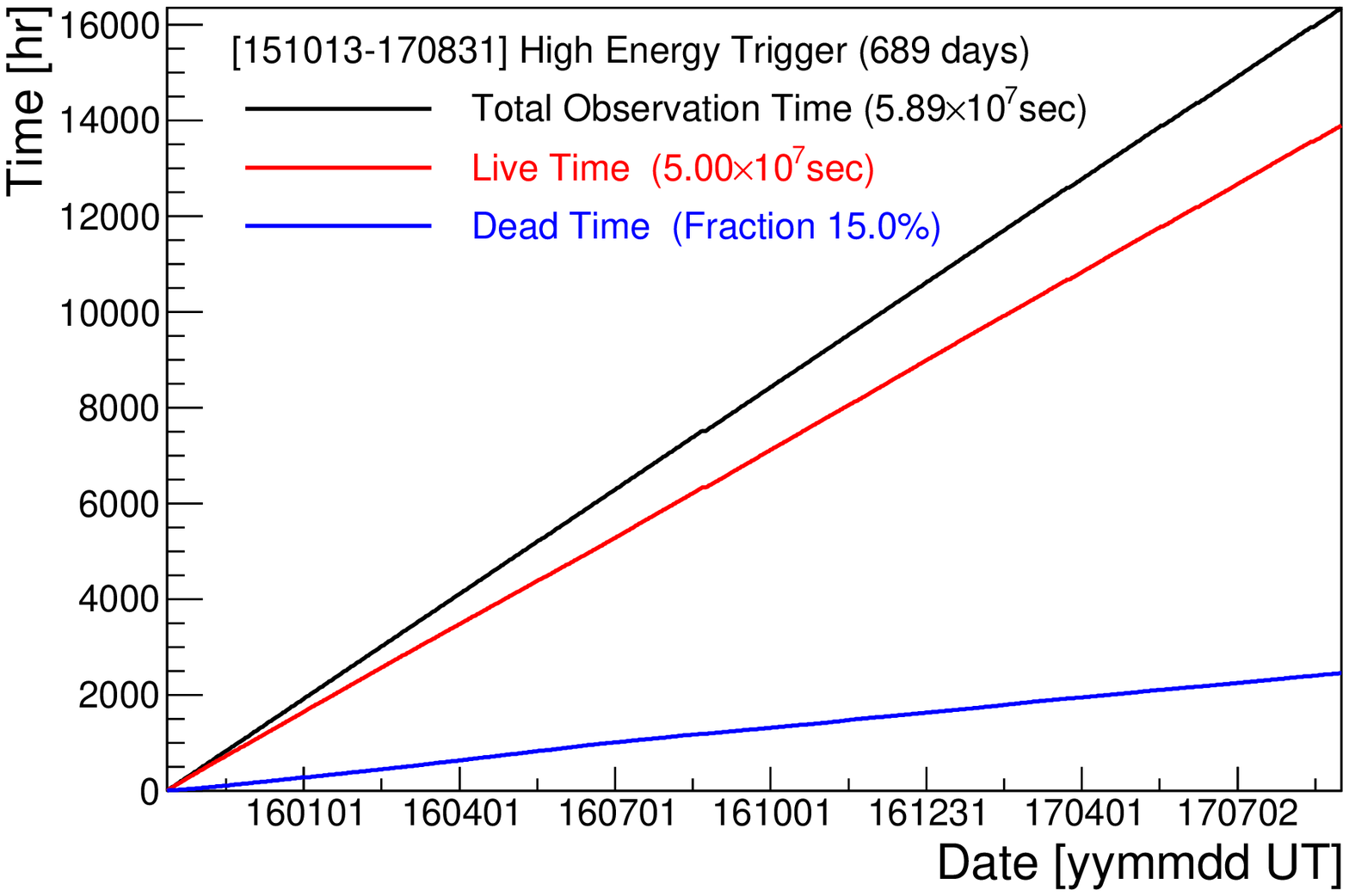}
\end{minipage}
\begin{minipage}{0.49\hsize}
\includegraphics[width=1.00\hsize]{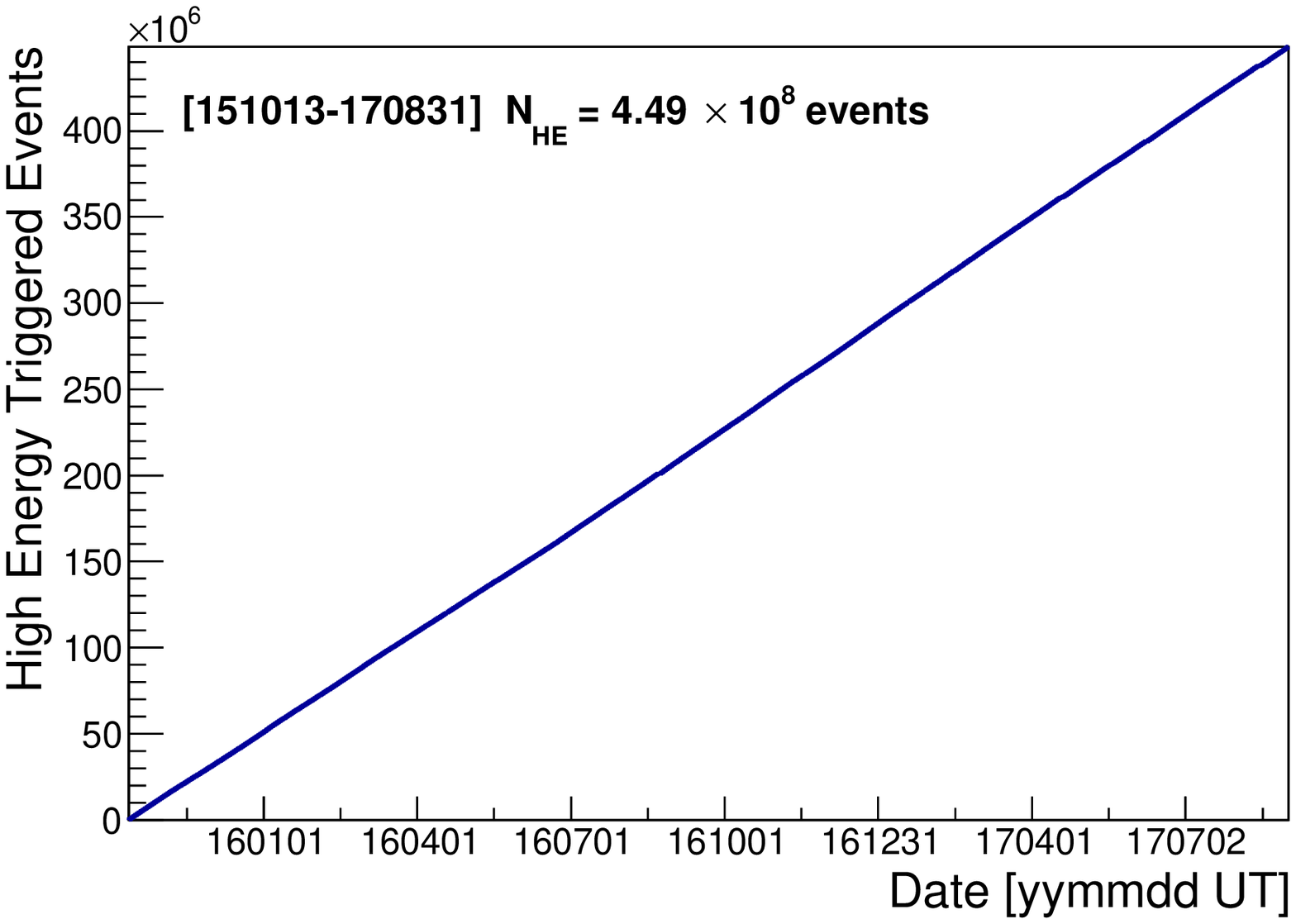}
\end{minipage}
\caption{({\it Left}) An accumulation of observation time (black line) for HE trigger. The red and blue lines indicate the live and dead times, respectively. The total live time reaches 5.89 $\times 10^7$ s.
({\it Right}) An accumulation of HE triggered events. The total number of HE events through the end of August, 2017 is 4.49 $\times 10^8$.} \label{fig:stat}
\end{figure*}

Figure \ref{fig:stat} shows for the HE-trigger the accumulated live time and number of events. 
Since October, 2015, the observation time has increased smoothly without serious trouble.
Transmission of data from JAXA-GSE to WCOC and processing of the data for scientific analysis at WCOC have also proceeded smoothly.
Furthermore, in order to monitor the daily performance of the CALET instrument, data quality check (DQC) plots are defined and processed every day. They are stored at WCOC and can be accessed 
through a website by the CALET science team members who take the shift at WCOC. 

\begin{figure*}[tbh!]
\begin{center}
\begin{minipage}{0.42\hsize}
\includegraphics[width=1.0\hsize]{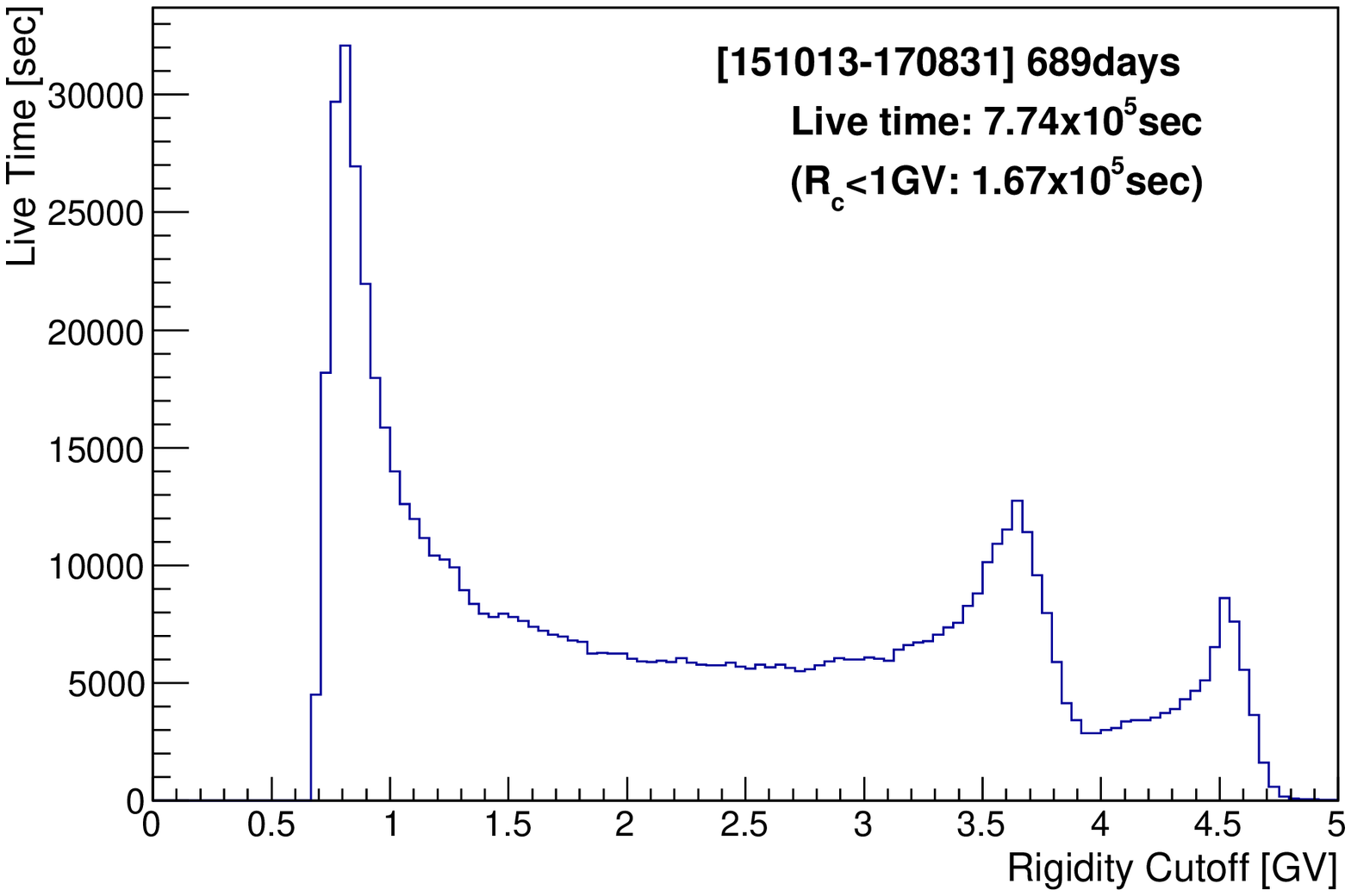}
\end{minipage}
\begin{minipage}{0.57\hsize}
\includegraphics[width=1.0\hsize]{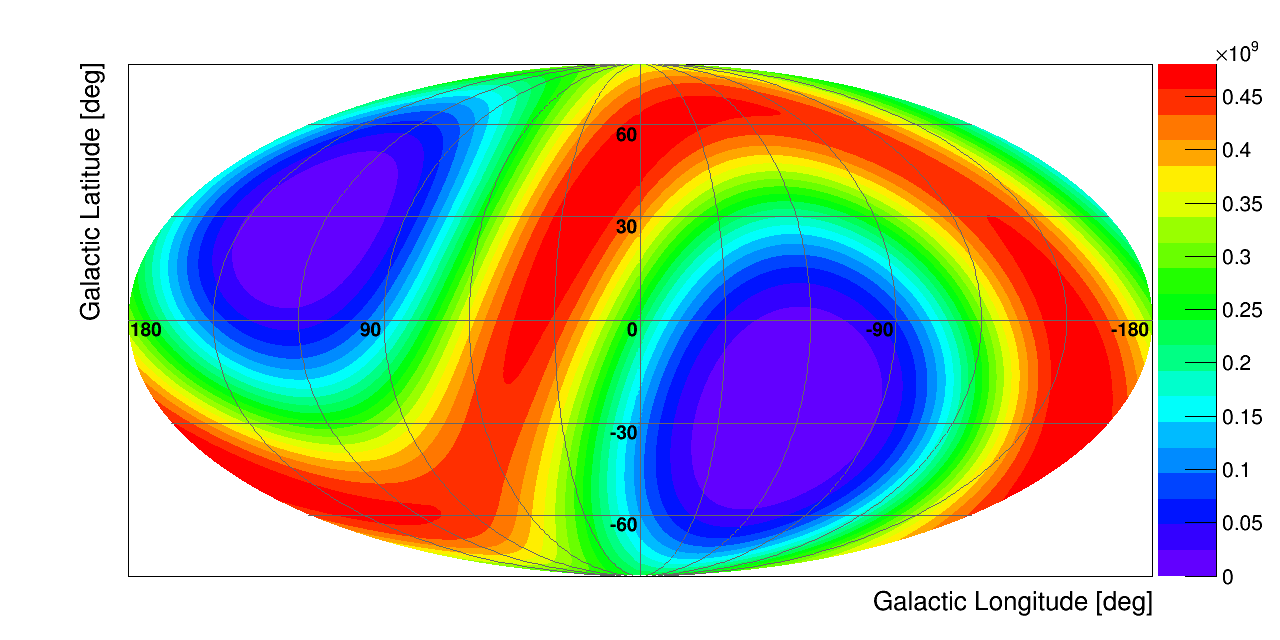}
\end{minipage}
\caption{({\it Left}) Rigidity cutoff distribution for the LE electron trigger.
({\it Right}) Effective exposure of the LE gamma-ray trigger at 3~GeV in unit of cm$^2$s through the end of August, 2017.
}\label{fig:le}
\end{center}
\end{figure*}
While the UH trigger is almost always active because of the low impact on the HE trigger,
the live times of LE triggers are limited.
As of the end of August, 2017, the total live time for the LE gamma-ray trigger is 8.42 $\times 10^{6}$~s, 
whereas that for the LE electron trigger is 7.74 $\times 10^{5}$~s. 
The rigidity cutoff distribution for the LE electron trigger is shown in the left-hand panel of Fig.~\ref{fig:le}. Since the LE-$e$ observation is performed in the high-latitude region, we have a significant
fraction of live time in the rigidity cutoff $<$1~GV. This makes it possible to measure
the temporal variation of the 1-GeV electron flux. 
The right-hand panel of Fig.~\ref{fig:le} shows the effective exposure of the LE gamma-ray trigger at 3~GeV in units of cm$^2$s. Since the LE-$\gamma$ observation is limited to low geomagnetic latitudes, exposure is not uniform in galactic coordinates. Note that the exposure for the HE trigger is much more uniform (not shown) because the observation mode is always active.

\section{Summary}
CALET has been delivering science data with stable instrument performance since October, 2015.
As of August 31, 2017, the total observation time is 689 days with a live time fraction close to 84\%. Nearly 450 million events have been collected with the high-energy ($E>$10~GeV) trigger.
In addition, a calibration data trigger mode, such as recording pedestal and penetrating particle events, a low-energy electron (gamma-ray) trigger mode operating at high (low) geomagnetic latitude, and other dedicated trigger modes are scheduled around the ISS orbit while maintaining the maximum exposure to high-energy electrons, gamma-rays, and nuclei.
The quality of the on-orbit data has been excellent and it is expected that a five-year CALET observation period will provide a wealth of new and interesting results~\citep{torii2015,pier2017,akaike2017,mori2017,nick2017,brian2017}.

\section*{Acknowledgments} 
We gratefully acknowledge JAXA's contributions to the development of CALET
and to operations on board the ISS. We also wish to express our sincere gratitude to ASI and NASA for their support of the CALET project. 
Finally, the work was supported in part by JSPS KAKENHI Grant Numbers JP26220708, JP17H02901, and by the MEXT Supported Program for the Strategic Research Foundation at Private Universities (2011-2015) (no. S1101021) at Waseda University.  
The CALET effort in the United States is supported by NASA through grants NNX16AB99G, NNX16AC02G, and NNH14ZDA001N-APRA-0075.

\section*{References}
\providecommand{\noopsort}[1]{}\providecommand{\singleletter}[1]{#1}%

\end{document}